\documentstyle[aaspp4,flushrt]{article}   	

\slugcomment{To appear in the Astronomical Journal, February 1998}

\def\ha{${\rm H\alpha}$}     

\def\HaNii{H$\alpha$+[\ion{N}{2}] $\lambda\lambda6548,84$}
\def\NII{[\ion{N}{2}] $\lambda\lambda6548,84$}

\def\oiii{[\ion{O}{3}] $\lambda5007$}

\newcommand{\HI}{\ion{H}{1}}

\newcommand{\kms}{km s$^{-1}$}

\newcommand{\whz}{W Hz$^{-1}$}
\newcommand{\ergs}{ergs s$^{-1}$}
\newcommand{\mJybeam}{mJy beam$^{-1}$}
\newcommand{\Msun}{{$M_\odot$}}
\newcommand{\Lsun}{{$L_\odot$}}

\lefthead{Morganti, Oosterloo, \& Tsvetanov}  
\righthead{Radio study of IC~5063}

\begin{document}

\title{A RADIO STUDY OF THE SEYFERT GALAXY IC~5063: \\ 
	EVIDENCE FOR FAST GAS OUTFLOW\footnote{Based on observations 
	with the Australia Telescope Compact Array (ATCA), which is 
	operated by the CSIRO Australia Telescope National Facility}}

\author{\sc R. Morganti} 
\affil{CSIRO, Australia Telescope National Facility, PO Box 76,
	Epping, NSW\,2121, Australia \\
	and Istituto di Radioastronomia, CNR, via Gobetti 101, 
	40129 Bologna, Italy}
\authoremail{rmorgant@atnf.csiro.au}

\author{\sc T. Oosterloo}
\affil{CSIRO, Australia Telescope National Facility, PO Box 76, 
	Epping, NSW 2121, Australia}
\authoremail{toosterl@atnf.csiro.au}
\author{\sc Z. Tsvetanov}
\affil{Department of Physics and Astronomy, Johns Hopkins University, 
	Baltimore, MD 21218, USA}
\authoremail{zlatan@pha.jhu.edu}

\medskip

\begin{abstract} 

We present new radio continuum (8 GHz and 1.4 GHz) and \HI\ 21 cm line
observations of the Seyfert 2 galaxy IC~5063 (PKS~2048-572) obtained
with the Australia Telescope Compact Array (ATCA).  The high
resolution 8~GHz image reveals a linear triple structure of $\sim 4''$
(1.5 kpc) in size.  This small-scale radio emission shows a strong
morphological association with the inner part of the optical emission
line region (NLR).  It is aligned with the inner dust lane and is
oriented perpendicular to the position angle of the optical
polarization.  We identify the radio nucleus to be the central blob of
the radio emission.  At 21 cm, very broad ($\sim 700$ \kms) \HI\
absorption is observed against the strong continuum source.  This
absorption is almost entirely blueshifted, indicating a fast net
outflow, but a faint and narrow redshifted component is also present.
In IC~5063 we see clear evidence, both morphological and kinematical,
for strong shocks resulting from the interaction between the radio
plasma and the interstellar medium in the central few kiloparsecs.
However, we estimate the energy flux in the radio plasma to be an
order of magnitude smaller than the energy flux emitted in emission
lines. Thus, although strong shocks associated with the jet-ISM
interaction occur and could contribute locally to the ionization of
the NLR, they are unlikely to account solely for the global ionization
of the emission line region, particularly at large distances.

The main structure of the \HI\ emission is a warped disk associated
with the system of dust lanes of $\sim 2'$ radius ($\sim$ 38 kpc,
corresponding to $\sim$ 5 effective radii).  The lack of kinematically
disturbed gas (both neutral and ionized) outside the central few kpc,
the warped structure of the large scale disk together with the close
morphological connection between the inner dust lanes and the
large-scale ionized gas, support the idea that the gas at large radii
is photoionized by the central region, while shadowing effects are
important in defining its X-shaped morphology.

From the kinematics of the ionized and of the neutral gas, we find
evidence for a dark halo in IC~5063, with very similar properties as
observed in some other early-type galaxies.

\end{abstract}


\section{INTRODUCTION}

The narrow-line regions (NLR) in Seyfert galaxies occupy the central
area in the host galaxy immediately surrounding the active nucleus.
Observationally, these are regions of highly ionized gas at radial
distances between $\sim 10$ pc (the $HST$ resolution at 10--15 Mpc) to
$\sim 1$ kpc from the nucleus.  The NLRs are kinematically complicated
and represent some of the best examples of regions where interaction
between the local ISM and radio plasma takes place.  The presence of
such interactions was first inferred from ground-based optical
spectroscopy of Seyfert galaxies with prominent linear radio sources. 
The emission lines in these regions usually have larger widths than
implied by simple gravitational motion and often show kinematical splits
as well as a morphological correspondence with the radio emission (e.g.,
Whittle et al.~1988).

In the past few years a number of Seyfert galaxies have been observed
by the $HST$ producing an impressive collection of images of their
NLRs (see Wilson 1997 for a review; Capetti et al.\ 1996 and
references therein).  In some objects, the association between the
radio plasma and the line emitting clouds is particularly striking.
The ionized gas often appears to form a `cocoon' around the radio
emission.  This supports the idea that the radio plasma compresses and
heats the ISM that, cooling down, produces the NLR clouds.  Shocks
may, therefore, be very important in determining the kinematics,
morphology and ionization state of the NLR.

In contrast, the extended emission line regions (EELR, Unger et al.\
1987) are traced up to tens of kpc from nucleus, and appear to have a
kinematically relaxed structure and no strong radio counterpart,
although their symmetry axis is invariably aligned with the position
angle (P.A.) of the elongated radio emission (Wilson \& Tsvetanov
1994).  The EELRs are believed to be mainly the result of
photoionization by the UV radiation from the active nucleus and, if
this is the case, the observed tight alignment implies that the radio
plasma and the ionizing photons are collimated by the same or by
strictly co-planar structures.

A careful comparison of the morphologies and kinematics of the gas in
different phases is a key element in understanding the physics of NLRs
in Seyfert galaxies.  In this respect, obtaining the parameters of the
neutral gas is of particular interest.  High resolution \HI\
observations give the distribution and kinematics of the {\sl cold}
component of the circumnuclear ISM and, therefore, complement the
optical data.  This has been done for a few objects.  For example, by
studying the neutral hydrogen in the center of NGC 4151, Pedlar et
al.\ (1992) found a close agreement between the kinematics of the
ionized gas and of the neutral hydrogen, strongly supporting the idea
that the ionized gas is simply a component of the gaseous disk,
ionized by an energy source in the nucleus.  This supports the
photoionization model and puts important constraints on the opening
angle of the ionizing radiation from the nucleus.  Additionaly, \HI\
absorption-line observations have the advantage of being able to
clearly distinguish between infall and outflow of gas and provide us
with information about the kinematics of the neutral gas along the
line of sight to the radio continuum.  In NGC 1068 and NGC 3079, high
resolution \HI\ observations (Gallimore et al.\ 1994) show evidence
for wind-driven outflow, indicating that in these galaxies shocks
could be important in determining the ionization structure of the gas.
Similar observations have been done by Brinks \& Mundell (1996).

To investigate these issues in more details, we have carried out a
study of the nearby Seyfert 2 galaxy IC~5063 (PKS~2048-572).  This
object has a number of interesting characteristics among which (1) the
Seyfert nucleus is hosted by an early-type galaxy, (2) it is
particularly strong in the radio continuum ($P_{\rm 1.4GHz} =
6.3\times 10^{23}$ \whz) allowing a comfortable study of the \HI\
absorption and (3) it is rich in \HI.

Despite these interesting characteristics, only a low resolution radio
image (made with the Fleurs radio telescope) and a total \HI\ profile
(obtained with the Parkes radio telescope) are available for IC~5063
(Danziger, Goss \& Wellington 1981, hereafter DGW81).  In this paper
we present new radio observations, both in the continuum and in the
\HI\ 21 cm line, obtained with the Australia Telescope Compact Array
(ATCA).  Our main goals are to map the morphology of the radio plasma
on the kiloparsec scale and compare it with that of the ionized gas,
and to study the kinematics of the neutral gas and compare it with
that of the ionized gas.

The paper is organized as follows: the basic properties of IC~5063 are
summarized in $\S$2 and the radio observations and the data reduction
are described in $\S$3.  In $\S$4 we present the major observational
findings and in $\S$5 we discuss their consequences for understanding
the physics of this complex system.  Throughout the paper we adopt a
Hubble constant of H$_0$ = 50 \kms\ Mpc$^{-1}$.

\section{BASIC PROPERTIES OF IC~5063}

IC~5063 is a nearby ($z=0.0110$) early-type galaxy hosting a Seyfert 2
nucleus and has been studied in different wavebands.  In the optical,
the surface brightness distribution is well described by a $R^{1/4}$
law and, therefore, it can be classified as an elliptical or S0
galaxy.  A gaseous disk is present and it has a complicated system of
(major axis) dust-lanes (DGW81; Colina, Sparks \& Macchetto 1991,
hereafter CSM91).  Moreover, the high contrast image of DGW81 shows
faint structures in the outermost regions reminiscent of tidal arms.
The optical spectra show strong emission lines (DGW81; Bergeron,
Durret \& Boksenberg 1983) with the ionized gas extending up to
$\sim$20 kpc and lying mainly in a disk (DGW81) but with anomalously
high velocities observed near the nucleus.  Bergeron et al.\ (1983)
and CSM91 detected a faint, very broad emission-line component and
Wagner \& Appenzeller (1989) found an off-nuclear region with broad
emission lines.

In polarized light IC~5063 shows high polarization in the near IR
(Hough et al.\ 1987) and a strong, broad \ha\ emission (Inglis et al.\
1993).  Like in some other Seyfert 2 galaxies this suggest that there
is a broad-line region which is obscured from our direct view and the
broad-line radiation is scattered into our line of sight by scatterers
outside the obscuring regions.  This is also suggested by the
detection of hard X-ray emission (Koyama et al.\ 1992) viewed through
a high column density ($N_{\rm H} = 2 \times 10^{23}$ cm$^{-2}$) and
with a X-ray luminosity and a spectral index in the range typically
found for Seyfert 1 galaxies.

IC~5063 is a strong $IRAS$ source (see Table~1) and has a warm
far-infrared excess peaking at 60 $\mu$m (Heisler \& Vader 1995).  It
has also been observed in CO by Wiklind et al.\ (1995) where it shows
a narrow profile ($\Delta V_{\rm CO}=163$ \kms).

Finally, the radio luminosity of IC~5063 is nearly two orders of
magnitude larger than typical for nearby Seyferts (Wilson 1991),
making it one of the strongest radio sources found in Seyfert
galaxies.  Because of this, it was suggested by CSM91 that this object
may well represent an intermediate or transition type between Seyfert
and radio galaxies.  Also, its \HI\ content is very high: DGW81 found
$1.0 \times 10^{10}$ \Msun\ that gives $M_{\rm HI}/L_{B}=0.19$, quite
anomalous for such a type of object (Wardle \& Knapp 1986).

The properties of IC~5063 are summarized in Table~1.

\section {OBSERVATIONS}

\subsection{Radio continuum observations}

IC~5063 has been observed with ATCA at 8~GHz to complement an optical
narrow-band imaging survey of a distance limited sample of southern
Seyfert galaxies (Tsvetanov, Fosbury \& Tadhunter in prep., hereafter
TFT; Morganti et al.\ in prep.).  The observations were done in July
1995 using the 6 km configuration (the longest available with ATCA).
We took data simultaneously at 8.256 and 8.896 GHz using a bandwidth
of 128 MHz for each of these frequencies.  These separate frequencies
allowed us to improve (radially) the {\sl uv} coverage.  This is
important in the case of ATCA because it is a six-telescope east-west
array.  We observed IC~5063 for about 6 hours in cuts of 15 min spread
over $2 \times 12$ hours.  The data reduction was done with the MIRIAD
package (Sault et al.\ 1995) which is particularly well suited for
ATCA data.

The 8~GHz image has a rms noise of 0.15 \mJybeam\ with a beam shape of
$1\farcs1 \times 0\farcs8$ elongated in position angle 51.8$^{\circ}$.
The resolution of this radio image is the highest reachable with ATCA
and matches the resolution of the optical narrow-band images.

An image of the radio continuum was also made at 1.4~GHz by using the
line-free channels in the \HI\ line observations (see below) and
combining the data from the 6 km (6D) and 750 m configurations (see
Table~2).  We obtained a rms noise of 0.8 mJy beam$^{-1}$ and a beam
(using uniform weighting) of $8\farcs2 \times 6\farcs7$
(P.A.~$-69^\circ$).

\subsection{{\sc H i} Observations}

As mentioned earlier, the aim of our \HI\ observations is to
investigate the properties of the neutral gas and compare them to
those of the ionized gas.  Given the extent of the ionized gas in
IC~5063, we have mainly used the longer configurations available with
the ATCA (6 km), but data with shorter configurations were also
collected to improve the sensitivity for low surface-brightness
extended emission.  We have made images of the line data using both
natural weighting (to study the overall distribution and kinematics of
the \HI) and uniform weighting (to study the \HI\ near the center).
Nevertheless, the final resolution of our \HI\ data is still much
lower than that of the optical data or of the 8 GHz radio continuum.
The data reduction was done using MIRIAD and GIPSY (Allen, Ekers \&
Terlouw 1985).  The configurations used as well as other observational
parameters are summarized in Tables 2 and 3.

An interference spike generated by the data acquisition system is
present at 1408~MHz, corresponding to a velocity of $\sim$2644 \kms\
and therefore at the edge of the \HI\ absorption profile of IC~5063
(see below). Although it limits slightly our ability to determine the
exact extent of the absorption profile, it does not in any way affect
the conclusions.

\subsection{Optical imaging}

IC~5063 was observed as part of a survey of volume limited sample of
southern Seyfert galaxies (TFT). All images of IC~5063 were taken with
the European Southern Observatory (ESO) Faint Object Spectrograph and
Camera 1 (EFOSC1) attached to the 3.6 m telescope at La Silla, Chile,
on 1993, July 14 and 15. The detector used was a Tek 512 CCD with
$\sim7$ electrons read-out noise. The pixel size of 27 $\mu$ provided
a scale of 0\farcs6075 pixel$^{-1}$ and just over 5$'$ field of view.

The galaxy was imaged through filters isolating the redshifted
positions of \oiii\ and \HaNii\ emission lines (on-band) and their
adjacent continua (off-band). The filters used have width of
$\Delta\lambda\sim$60\AA\ and $\sim$70--80\AA\ in the \oiii\ and
H$\alpha$ wavelength region, respectively. In the case of IC~5063 we
took two exposures of 10 min per filter to allow for cosmic-ray hits
cleaning afterwards.

The images were processed with the IRAF\footnote{The Image Reduction
and Analysis Facility (IRAF) is distributed by the National Optical
Astronomy Observatories, which is operated by the Association of
Universities for Research in Astronomy, Inc., under contract to the
National Science Foundation} software package. Reduction steps
included: CCD bias subtraction, flat-fielding, and sky
subtraction. The images were then registered using several well
exposed stars in the field, and frames through the same filter were
combined to clean the cosmic ray events and improve the
signal-to-noise ratio. Small differences in the seeing were eliminated
by convolving with a Gaussian to match the point spread functions
(PSF). The final resolution of the of the IC~5063 images is 1\farcs7
(FWHM).

Emission line maps were formed by scaling and subtracting the
continuum images from the line+continuum ones. The scaling factors
were determined from the emission-line free regions in the galaxiy.
Finally, the images were flux calibrated using the observations of
spectrophotometric standard star through the same filters taken during
the same night.  We estimate that the overall flux calibration of the
emission line maps is accurate to within 10\%--15\%. The noise level
of individual frames is of order $10^{-17}$ ergs cm$^{-2}$ s$^{-1}$
px$^{-1}$.



\section{RESULTS}

\subsection{The radio continuum}

The 8~GHz radio continuum image is shown in Fig.~\ref{Fig:3cm}.  This
image reveals a linear radio structure consisting of three clumps
oriented in P.A.\ $\sim 295^\circ$.  This morphology is common in
Seyfert galaxies (Ulvestad \& Wilson 1984, 1989).  The total flux
density at 8~GHz is 230 mJy ($\log P_{\rm 8 GHz}$ = 23.11 \whz) and
the extent is about 4$''$, corresponding to 1.3 kpc.  Most of the flux
(195 mJy) comes from the NW blob which clearly dominates the emission
at that frequency.  The central blob has a flux density of 16 mJy,
while the eastern one is only 9 mJy.  Despite the relatively high
observing frequency, the blobs are unpolarized (we measure a
fractional polarization of 0.5\% only in the brightest blob).  The low
polarization is common in Seyfert galaxies, likely due to the Faraday
depolarization from the dense gas around (Wilson 1991).

The line-free channels of the \HI\ data have been used to make an
image of the continuum emission at 1.4 GHz.  The final map is
presented in Fig.~\ref{Fig:21cm-opt} superimposed onto the optical
image from the Digitized Sky Survey (DSS) where the outer dust lane in
IC~5063 is clearly visible.  At 1.4 GHz we find a flux density of 1.26
Jy, in good agreement with the flux density derived by DGW81.  Thus,
our flux density measurement at 1.4~GHz (together with the data from
literature), confirms the steep value of the overall spectral index
($\alpha \sim -1.1$ for $ S \propto \nu^\alpha$) of IC~5063.  Given
that we have only one frequency (8~GHz) with sufficient resolution to
separate the three blobs, we cannot estimate their spectral indices
separately.  The 21 cm radio continuum position is very close to that
of the brightest 8 GHz blob (shifted slightly, by about 0\farcs5 $\pm$
0.04, toward the east, see Fig.~\ref{Fig:3cm}), indicating that also
at 21 cm most of the flux comes from this region.

At 1.4~GHz the radio continuum is extended in a direction
perpendicular to that at 8~GHz and it is roughly aligned with the
minor axis of the galaxy and perpendicular to the dust lane. This
emission, however, could be of different origin.  Baum et al.\ (1993)
have observed that the {\sl large scale} radio structure in Seyfert
galaxies is often aligned with the minor axis of the galaxy, and that
this radio emission could be related to a starburst driven wind. They
also find that in Seyfert galaxies the {\sl small scale} radio
structure in general has no correlation with either the large scale
radio structure or with the optical orientation of the galaxy. It
appears that IC~5063 behaves in a similar way.  However, in active
galactic nuclei hosted by early-type galaxies the radio emission is
usually perpendicular to the dust lane (Kotany \& Ekers 1979,
M\"ollenhoff, Hummel \& Bender 1992, van Dokkum \& Franx 1995). This
has been used to support the idea that the gas could fuel the activity
in an active nucleus (Kotany \& Ekers 1979), although a number of
exceptions exist (M\"ollenhoff, Hummel \& Bender 1992).

Finally, the position angle of the inner radio structure appears to be
approximately perpendicular to that of the optical polarization
(P.A.~34$^{\circ}$, Inglis et al.\ 1993) as expected if the optical
emission (in this case from the broad-line region) is scattered into
the line of sight by electrons and/or dust grains.  This is therefore
in agreement with the prediction from the unified schemes.

The comparison between the radio and the ionized gas is of interest.
The new optical narrow-band images (TFT) confirm the original result
of CSM91 that the high-ionization line-emitting gas has an
``X-shaped'' morphology and show that the ionized gas can be traced
out to a distance larger than $\sim 30''$.  There are three radial
filaments of highly ionized gas elongated along P.A.\
285$^{\circ}-290^{\circ}$, P.A.\ 310$^{\circ}-315^{\circ}$, and P.A.\
325$^{\circ}-330^{\circ}$. The basic symmetry axis of these extended
emission line regions is roughly coincident with the major axis of the
light distribution (P.A.\ $\sim 303^{\circ}$).  The opening angle of
the X-shaped structure is $\sim 50^{\circ}$.

The overlay of the 8~GHz map to the \oiii\ image from TFT is presented
in Fig.~\ref{Fig:3cm-oiii}.  The radio emission is closely aligned
with the symmetry axis of the ionized gas.  For a sample of Seyfert
galaxies, Wilson \& Tsvetanov (1994) have found a tight alignment
between the ``ionization cone'' and the radio axis.  This seems to be
the case also in IC~5063, although the ionized gas is in a X-shaped
structure more than in a cone.  Moreover, if we concentrate on the
inner optical region we can see that the radio and optical structures
are very similar as shown in the enlargement in
Fig.~\ref{Fig:3cm-oiii}.  This is not really surprising given that it
is well known that the NLR (i.e.\ ionized gas on the scale of few kpc)
is invariably co-spatial with the radio emission (Wilson \& Ulvestad
1983; Haniff, Wilson \& Ward 1988).

An $HST$ image of IC~5063 has been obtained from the the public
archive.  This image is a single 500~s exposure obtained with WFPC2
through filter F606W and shows even more impressively the complicated
structure of the dust lane in the central region (see
Fig.~\ref{Fig:ratio-hst} insert).  Already from the ground-based
optical images, CSM91 described the dust lanes in IC~5063 as a
``complicated zig-zagging distribution running approximately parallel
to the major axis and mainly concentrated in the northern side and
most symmetric nearest the nucleus''. As we discuss briefly in
$\S$5.4, the observed morphology can probably be explained as the
result of a warped structure.

The insert in Fig.~\ref{Fig:ratio-hst} shows the radio map at 8~GHz
superposed onto the $HST$ image.  Given that the F606W filter includes
bright emission lines (\oiii, \ha\ and \NII), the observed morphology
is strongly influenced by the emission-line gas.  The overlay shows
that the central radio blob corresponds to the peak of the light
distribution. Although there is, as always, some uncertainty in the
alignment of these images, the error in the relative positioning is at
most 0\farcs5. Comparing the 8~GHz image with the ground based \oiii\
image as well as the ground based \oiii\ image with the $HST$ image,
we can exclude that the brightest radio blob corresponds to the center
of the galaxy.  More difficult, also because of the low resolution of
our radio image, is to say if the optical emission really ``wraps
around'' the radio blobs like in a number of other Seyfert galaxies
that have been studied at high resolution (e.g. Mrk~3, Capetti et al.\
1995), although data suggests that also in IC~5063 this may be the
case.  We have obtained higher resolution VLBI radio data to
investigate the small scale structure of the radio emission and to
attempt a better comparison with the structure of the ionized gas on
the scale of the $HST$ observations. Results from the VLBI
observations will be presented in a future paper.

\subsection{The Neutral Gas}

In our study of the neutral gas in IC~5063 we detect the \HI\ 21 cm
line in absorption as well as in emission. These correspond to two
very different components, both kinematically and spatially, of the
local ISM and we discuss them separately below.

\subsubsection{The {\sc H i} absorption}

The most striking feature in the \HI\ data is the presence of very
wide, mostly blueshifted absorption against the central continuum
source.  The position-velocity map taken along the major axis
(Fig.~\ref{Fig:xv120}) outlines the individual features of this broad
absorption particularly well.  The absorption profile is very broad
($\sim 700$ \kms, because of the interference at 1408 MHz,
corresponding to 2644 \kms, the accuracy of this determination is
somewhat compromized), and almost entirely blue shifted, far beyond
the velocities characteristic of the \HI\ emission.  The large
velocity range and the asymmetry makes it unlikely that the motion is
gravitational.  Instead, it is more likely that we see a strong net
outflow.  The shape of the absorption profile
(Fig.~\ref{Fig:abs-prof}) suggests that there are several \HI\ clouds
along the line of sight, the main clouds being roughly 600 \kms\ and
150 \kms\ from the systemic velocity.  A faint redshifted component is
also visible.

It is important to remember that even in the 21 cm images made with
uniform weighting, the achieved resolution is FWHM $\sim 7''$ and the
central continuum source is only slightly resolved.  Therefore, we
cannot make a detailed determination against which of the three blobs
observed at high frequency the absorption is occurring.  However,
averaging the line channels containing only \HI\ absorption of the
uniformly weighted data, the position of the absorption appears to be
offset by 0\farcs5 W with respect to the peak of the continuum at 21
cm and appears to be coincident with the position of the brightest
blob in the 8~GHz map (see Fig.~\ref{Fig:3cm}). This would mean that
{\sl at least the very blueshifted absorption is not against the
nucleus but against the W blob}.  We have estimated the level of
significance of this apparent offset by means of a Monte Carlo
simulation. We have made a large number of model images containing a
point source of the appropriate strength and noise level. Fits to
these model images indicate that the probability that the observed
offset is due to noise effects is 4\%.

This offset is in agreement with results from optical observations. In
the region $\sim 2''$ NW from the center of the galaxy, i.e.,
co-spatial with the brightest radio knot at 8 GHz, Wagner \&
Appenzeller (1989) found unusually broad wings (several hundred \kms)
of the permitted as well as forbidden lines.  The peak of these broad
profiles appears blueshifted by $\sim$ 150 \kms\ with respect to the
systemic velocity, with wings out to velocities similar to what we
observe in the \HI\ absorption.  Wagner \& Appenzeller (1989)
interpreted these broad lines as nuclear emission scattered into the
line of sight, although they could not rule out gas outflow. However,
given the \HI\ absorption we observe, gas outflow becomes the more
likely hypothesis to also explain the optical observations.  This
strongly suggests that in the region of the brightest radio blob at 8
GHz a strong interaction between the radio plasma and the ISM takes 
place.

In the channels where there is both absorption and emission, the
position of the absorption shifts towards the peak of the 21 cm
continuum, but this is (at least partially, perhaps completely) due to
confusion with the \HI\ emission.

\subsubsection{The {\sc H i} emission}

IC~5063 contains an unusually large amount of \HI\ for a galaxy of its
type.  We estimate $8.4 \times 10^{9}$ \Msun\ of \HI\ over a wide
velocity range: from $\sim$3150 to $\sim$3650 \kms.  The global \HI\
profile is shown in Fig.~\ref{Fig:prof-hi}.  The difference with the
value obtained from Parkes observations ($10^{10}$ \Msun, DGW81) is
very small and likely due to the relatively high resolution of our
observation.  Because the aim of our \HI\ observations is to
investigate the kinematics of the neutral gas and compare it with that
of the ionized gas, the combination of arrays used in this work is not
the most suitable for mapping very extended low-surface brightness
\HI\ emission.  A proper study of the large scale characteristics of
the \HI\ emission is being carried out by Blank et al.\ (in prep).

The position-velocity map along the major axis (P.A.\ 120$^\circ$) is
shown in Fig.~\ref{Fig:xv120}.  This plot shows that, to first order,
the \HI\ is in a regularly rotating disk with systemic velocity of
$\sim 3400$ \kms.

Figure~\ref{Fig:chan} shows images from the individual velocity
channels (plotted every second channel).  Because of the strong
absorption, the maps of the total intensity and the intensity-weighted
mean velocity of the \HI\ emission were made in the following way.
The line maps were smoothed spatially to a resolution of 30 arcsec,
after the negative values corresponding to the absorption were set to
zero.  This smoothed cube was used to mask the original cube: pixels
with signal below 2$\sigma$ in the smoothed cube were set to zero in
the original cube.  The total intensity image and the velocity field
were derived from this masked version of the full resolution data.
Fig.~\ref{Fig:tothi-opt} shows the total \HI\ intensity map of IC~5063
superimposed on the optical image of the galaxy.  The ``hole'' in the
center of the \HI\ distribution (and coincident with the center of the
galaxy) represents the region of strong absorption against the radio
continuum.

The total \HI\ intensity image and the velocity field
(Fig.~\ref{Fig:vel}) show that the main structure of the \HI\ emission
is that of a disk, oriented in a direction very similar to the system
of dust lanes (as in the majority of the dust-lane ellipticals with
\HI, Morganti, Sadler \& Oosterloo 1997).  We find that the \HI\
emission extends to $\sim 2'$ radius, corresponding to just over 5
effective radii ($R_{\rm eff}$ = 22\farcs4; CSM91).  The pattern of
the dust lanes in IC 5063 already suggests that this galaxy contains a
warped disk (e.g.~DGW81).  A number of dust lanes are in fact evident,
each having a sightly different orientation, suggesting that IC~5063
is perhaps to some extent similar to galaxies like NGC~4753
(Steiman-Cameron, Kormendy \& Durinsen 1992) where there is strong
evidence for a gaseous disk precessing differentially in the potential
of the galaxy.  The \HI\ velocity field confirms that the disk in
IC~5063 is warped in a way one would expect from the morphology of the
dust-lanes. From Fig.~4b in CSM91 one can see that there are three
major dust-lanes: one close to the center, an intermediate dust-lane,
making an angle of about 20$^\circ$ with the inner dust-lane, and an
outer dust-lane that is more or less parallel with the inner one. The
change in kinematic position angle that is evident in the velocity
field corresponds to the change in orientation between the
intermediate and outer dust-lane. Also at other positions (e.g. $\sim
30''$ east of the center) the velocity field shows kinks in the
velocity contours, also characteristic of a warp.

\subsection{Nearby objects}

DGW81 reported one other radio source about 4$'$ west of IC~5063. This
object was also observed at 843~MHz by Jones \& McAdam (1992) and is
part of the Molonglo catalogue of radio sources (MRC 2047--572) and
appears to be unrelated to IC~5063.  MRC2047--572 is also visible in
our 1.4~GHz image (see Fig.~\ref{Fig:21cm-opt}).  From our data the
position of this radio source can be determined much more accurately,
and we now find marginal evidence for an optical identification for
this source.  The higher resolution of our data shows that the
extension observed in this source by both DGW81 and Jones \& McAdam
(1992) actually consists of two (background?) radio sources.  In
Fig.~\ref{Fig:21cm-opt} we show an overlay of the 1.4~GHz image on top
of the optical image taken from the Digitized Sky Survey, showing the
two radio sources just east of MRC~2047--572.  This figure also shows
that there is a background group or cluster of galaxies near IC~5063
with which these radio sources are possibly associated (as perhaps
also their morphology suggests).  From our data we obtained a flux of
320 mJy for MRC~2047--572 and $\sim$97 mJy for the two other sources
together.  Comparing this with the flux derived from MOST observations
(870 mJy, Jones \& McAdam 1992) that includes all these sources, we
obtain a steep overall spectral index of $\alpha \sim -1.4$.

We do not detect \HI\ emission from any other object in the field.

\section{DISCUSSION}

\subsection {The Strong Nuclear Outflow}

There is a growing body of evidence that nuclear gas outflows are
common in AGNs. In type 1 Seyfert galaxies, where our view of the
nucleus is relatively clear, these are detected through absorption on
the line of sight to the central UV and/or X-ray source. Recent $HST$
observations indicate that at least 50\% of Seyfert 1 galaxies have
intrinsic UV absorption lines (e.g. Grenshaw 1997). Despite the small
number of objects for which this data is available so far, a number of
common properties are emerging.  In all cases, the UV absorption is
blueshifted (up to $-$1500 \kms\ relative to systemic) indicating a
net outflow. The absorption profiles are often broad (a few hundred
\kms) and consist of multiple components suggesting macroscopic bulk
motions. Absorption lines from elements and ions covering a large
range of ionization stages have been detected. In the best studied
cases such as NGC 4151 (Kriss et al.\ 1992) and NGC 3516 (Kriss et
al.\ 1996), where spectra from the Hopkins Ultraviolet Telescope
($HUT$) and $HST$ are available, these include hydrogen Lyman series,
and ionization stages as high as \ion{O}{6}.  A number of attempts
have been made to relate the UV absorbers to the warm absorbers (cf
Nandra \& Pounds 1994) seen in the X-ray spectra of many Seyfert
galaxies (Mathur 1994; Mathur et al.\ 1995), but it is clear that the
single zone model is an oversimplification (e.g. NGC 3516; Kriss et
al.\ 1996).

In Seyfert 2 galaxies the nucleus is obscured at all wavelengths from
the near IR to the soft X-rays, and the evidence for outflows is
mostly indirect. The outflows are inferred mainly from the detection
of asymmetric wings and/or splitting of the lines from the resolved
emission-line structure often associated with the radio emission
(Whittle 1989; Unger et al.\ 1987). More recently, detailed multislit
spectroscopy (e.g.~Mulchaey et al. 1992) and/or integral field
spectroscopy (Arribas et al.\ 1997 and references therein) of EELR in
nearby Seyferts have yielded similar results. A number of good
examples are summarized in Aoki et al.\ (1996).

Mapping the \HI\ velocity profiles offers an opportunity to detect
outflows unambiguously in Seyfert 2's and to compare the kinematics of
neutral and ionized fractions in Seyfert 1's. The striking result
about IC~5063 is that such broad absorption does not seem to be at all
common in Seyfert 2 galaxies, with the exception of NGC~1068.  Clearly
this broad absorption is the result of an interaction of a nuclear
outflow with the gas in the circumnuclear ISM. However, the picture is
further complicated by the fact that at least the most blueshifted
absorption in IC~5063 is not against the nucleus but against the
western radio blob (see $\S$4.3 and Fig.~\ref{Fig:3cm}).  The
situation is not like in NGC 1068 (Gallimore et al.\ 1994) where the
more blueshifted absorption is seen against the core while the
absorption against the jet is less blueshifted. Clearly, VLBI
observations are needed to confirm our findings, and they will likely
have to be combined with high resolution UV ($HST$) and X-ray
observations to try to unravel this complex kinematics.

The \HI\ column density we derive from the absorption profile is
$N_{\rm HI} = 1.0 \times 10^{21}$ atom cm$^{-2}$, assuming a spin
temperature $T_{\rm spin}=100$ K.  This value of $N_{\rm HI}$ is very
similar to the column densities found in other Seyfert galaxies (also
assuming $T_{\rm spin}=100$ K): ($3.9\pm0.5) \times 10^{21}$ atom
cm$^{-2}$ for NGC~4151 (Mundell et al.\ 1995), and $(1-4) \times
10^{21}$ atom cm$^{-2}$ for NGC~1068 (Gallimore et al.\ 1994).
However, the column density we derive is likely to be a lower limit
for several reasons.  First, given the relatively low resolution, the
absorption is likely to be confused by some \HI\ emission from the
disk.  Second, in estimating the optical depth, we have assumed that
the absorption is uniform over the continuum source.  It is, however,
possible that the absorption is seen against only one or two of the
blobs observed at 8 GHz, and perhaps we even see evidence for that
(see $\S$4.3).  This would lead to underestimating the optical depth.
Let us for the sake of argument assume that the structure in the
continuum at 21 cm on an arcsecond scale consists also of 3 blobs.
Then if all the absorption is against the brightest radio source, the
above column density would not increase significantly, since most of
the 21 cm flux comes from that blob.  In the extreme case where the
absorption is against the core, the estimated column density would
depend strongly on the spectral index one wants to assign to the core.
If we assume a flat spectral spectral index for the core, the optical
depth becomes very high given that the core flux (16 mJy) would be the
same as the strength of the absorption.  In the case of a steep
spectral index (e.g., $\alpha=-1$) the core flux would be $\sim$100
mJy and the column density would become $1.3 \times 10^{22}$ atoms
cm$^{-2}$.  Only higher resolution observations could clarify this
point.

Finally, and probably most importantly, the presence of a strong
continuum source near the \HI\ gas may significantly increase the spin
temperature because in this case the radiative excitation of the \HI\
hyperfine state can dominate the usually more important collisional
excitation (e.g., Bahcall \& Ekers 1969).  We have estimated the
magnitude of this effect and find for a wide range in volume densities
of the \HI\, that if the absorbing \HI\ gas is closer than about 500
pc from the strongest radio source, the spin temperature is at least
several thousand Kelvin. So it is quite likely that the column density
corresponding to the \HI\ absorption is at least $10^{22}$ cm$^{-2}$,
while it is not even impossible that it is as high as $\sim 10^{23}$
cm$^{-2}$, similar to what is derived from X-ray observations (Koyama
et al.~1992).

The redshifted component of the \HI\ absorption might be associated
with a nuclear torus/disk.  The width of the CO profile as observed by
Wiklind et al.\ (1995) is very narrow compared to the total
\HI-emission profile.  This clearly means that the CO emission does
not come from the molecular counterpart of the \HI\ disk.  Assuming
that the range of velocities in the CO is due to rotation, the
inclination of the CO disk must be quite different from that of the
\HI\ disk.  Interestingly, the half-width of the CO profile (82 \kms)
corresponds quite closely to the width of the redshifted absorption
component, suggesting perhaps that both originate in the same
structure.

The blueshifted absorption observed in IC~5063 is also very unusual
when compared to other nearby elliptical galaxies (of which none have
a Seyfert nucleus) that have nuclear sources of comparable radio
power.  In these nearby elliptical galaxies, if absorption is observed
against the central continuum source, this absorption is invariably
redshifted and is much narrower in velocity, as found by van Gorkom et
al.\ (1989).  Narrow absorption has also been detected in a number of
more powerful radio galaxies (Cygnus~A, Conway \& Blanco 1995;
Hydra~A, Dwarakanath, Owen \& van Gorkom 1995 and NGC 4261, Jaffe \&
McNamara 1994).  The fact that, on the contrary, IC~5063 shows the
broad blueshifted absorption could be due to some stronger kind of
interaction going on in this object, possibly because of the richer
ISM.  This could be related to the fact that the radio jet is aligned
with the dust-lane, i.e.~it escapes the nucleus in the direction of
the disk of gas and therefore the interaction between the radio plasma
and the environment is particularly strong.

\subsection {Shocks in the NLR}

The radio source in IC~5063 is unusually strong when compared to those
observed in typical Seyfert galaxies.  Nevertheless, IC~5063 exhibits
characteristics similar to those of other Seyfert galaxies studied at
radio wavelengths.  In particular, the linear radio structure is only
a few kpc in size, it is coincident with the inner, bright, part of
the optical emission line region, and its position angle is very close
to the symmetry axis of the extended emission line region.

The comparison of the 8 GHz radio map with the $HST$ image
(Fig.~\ref{Fig:ratio-hst}) reveals morphological characteristics
similar to what is observed in other Seyfert galaxies with close
association between radio and line emission.  Although the uncertainty
in the alignment of the radio and $HST$ images does not warrant a
detailed investigation, the optical emission could be ``wrapping
around'' the radio blobs as found in cases like Mrk~3 (Capetti et al.\
1995). Because of this close association, the structure of the line
emitting gas in the NLR of IC~5063 is likely to be dominated by the
compression and heating of the interstellar gas generated by shock
waves produced by the supersonic radio plasma.

The presence of strong outflow in IC 5063 also suggests strongly that
fast shocks driven by the radio plasma must be present in the
circumnuclear region.  This raises the usual question of the
importance of shocks as ionization mechanism in Seyfert galaxies.
Models involving bow shocks driven into the circumnuclear medium by
radio jets have been proposed by Pedlar, Dyson \& Unger (1985) and
Taylor, Dyson \& Axon (1992).  These models generate high velocities
and large velocity dispersions in the line emitting gas, but continue
to rely on UV radiation from the nucleus to ionize the gas compressed
by the shocks.  More recently, Dopita \& Sutherland (1995, 1996) have
suggested that radiation from the shocks themselves plays the dominant
role in the ionization of the line emitting gas.

We have used energy budget arguments to investigate if the energy
supplied by the radio plasma is sufficient to explain the energy
radiated in the emission lines.  Following the calculations presented
by Bicknell (1995), we find that the energy flux derived from the
radio (using our 8~GHz data) is $\sim 2 \times 10^{42}$ \ergs.  The
observed \oiii\ luminosity is $L_{\rm [OIII]}=6.6 \times 10^{41}$
\ergs\ (CSM91) and the great majority of this is coming from the
central region coincident with the radio emitting region.  The total
line luminosity can be estimated to be about 10--15 times the \oiii\
luminosity.  Clearly, unless the radio lobes are well out of
equipartition or the lobe plasma contains a significant thermal
component (Bicknell et al.\ 1998), the radio energy is not sufficient
to completely power the line emission. As the broad \HI\ absorption
shows, fast shocks do occur in IC~5063 and they could be important for
the ionization locally (similar to as was shown to be the case in
NGC~1068; Capetti et al. 1997), but it seems that for the ionization
of the NLR of IC~5063 the UV photons from the active nucleus are also
required.  This appears to be observed in basically all Seyfert
galaxies (Bicknell et al.\ 1998).

We also note that the brightness distribution of the optical emission
line gas in the central few kpc is quite symmetrical with respect to
the center, while the radio emission is very asymmetric.  This also
suggests that the interaction between the radio plasma and the
surrounding gas is not the only mechanism responsible for the
ionization of the NLR.

\subsection {Shadowing and ionization from nucleus}

The other interesting aspect of our observations is the relationship
between the ionized gas and the neutral gas we see in emission.
Outside the central region where the absorption occurs, the kinematics
of the \HI\ is very regular and there are no indications for
interaction between the radio plasma and the \HI\ disk.  Thus, the
interaction between radio and ISM appears to be confined to the
central region.  Given that the \HI\ appears to fill the disk rather
well, the sharp and straight edged structure of the ionized gas must
be connected with the shape of the radiation field (typical of
photoionization, Morse et al.\ 1996).  A comparison of the morphology
of the highly ionized gas (e.g., ratio \oiii/\ha) with that of the
inner dust structures visible in the $HST$ image, combined with the
fact that the gas disk in IC~5063 is warped, reinforces the suggestion
of CSM91 that shadowing effects are important and that the gas at
larger radii is photoionized by the central UV source.  This is
illustrated in Fig.~\ref{Fig:ratio-hst} where we have overplotted the
distribution of the high-excitation ionized gas (as derived from the
\oiii/\ha\ ratio map from TFT) onto the $HST$ image.  As noted before
by CSM91, the location of the inner dust lane is very suggestive for
the morphology of the high-excitation gas being determined by
shadowing effects caused by the inner dust lane.

Figure~\ref{Fig:tothi-oiii} shows the total \HI\ superimposed to the
\oiii\ image.  The extent of the ionized gas coincides very well with
that of the \HI\ disk.  The \HI\ velocities near the center (see the
position-velocity map in Fig.~\ref{Fig:xv120}) also correspond exactly
to the velocities of the ionized gas measured by DGW81 and CSM91 along
the major axis of the ionized gas distribution.  The neutral gas does
not appear to be disturbed in the region outside the inner few kpc.
This suggests that in IC~5063 we are observing a situation similar to
NGC~5252 (Prieto \& Freudling 1993, 1996).  Based on images of the
\HI, combined with narrow-band images and kinematical information of
the ionized gas, Prieto \& Freudling concluded that in NGC~5252 there
is single gas system which is only partly ionized because of the
anisotropy of the ionizing radiation.  The fact that in IC~5063 the
ionized gas has the same extent as the \HI, and the kinematics of the
two appear to connect smoothly, suggest that also in this object the
neutral and the ionized gas are physically connected.  For NGC~5252,
Prieto \& Freudling (1996) argued that the fact that both physical
states can exist at a similar distance from the nucleus implies that
the radiation field must be anisotropic.  It appears that this
argument also applies to IC~5063.

\subsection {The {\sc H i} disk and its origin}

The \HI\ emission is extended in the direction of the dust-lane.  This
has been found in most of the dust-lane ellipticals that show \HI\
emission (Morganti, Sadler \& Oosterloo 1997; also NGC~1052, van
Gorkom et al.~1986, and Centaurus~A, van Gorkom et al.~1990).

As for all the early-type galaxies with a detectable amount of \HI,
the origin of the neutral gas in IC~5063 is believed to be external
due to a merger with a gas-rich galaxy.  The merger hypothesis for
IC~5063 has been already suggested both by DGW81 and CSM91 for a
number of reasons.  For example, the fact that dust lanes appear to be
more symmetric near the center supports the idea of an external origin
for the dust.  As in the case of e.g.~NGC~5266 (Morganti et al.\ 1997)
the amount of \HI\ present in IC~5063 is atypical for a galaxy of its
type, the observed $M_{\rm HI}/L_B$ is more characteristic of a
late-type spiral. Therefore, the origin of the \HI\ in IC~5063 is very
likely a merger between two (or more) spiral galaxies, not just an
accretion of a small gas-rich object.  The effect of this event can be
seen also from the faint optical ``tails'' observed by DGW81 in their
deep optical image.  Our data do not have the right combination of
resolution and sensitivity to find a possible low-surface-brightness
\HI\ counterpart of these faint optical tails, as is observed in
e.g.~NGC~5266 (Morganti et al.~1997) or in some shell galaxies
(Schiminovich et al.~1995)

Most of the neutral gas appears to be in a disk, suggesting that the
gas is settled and the merger is relatively old.  Because of this, we
can make an estimate of the $M/L$ for IC~5063 at large radius.  We
estimate the projected rotation velocity to be about 240 \kms.
Assuming a spherical mass distribution, this gives a $M/L_B \sim 14$
at 5.4 $R_{\rm eff}$, where we have assumed an inclination of
$74^\circ$ as determined from the axial ratio of the \HI\ disk.  The
optical data of DGW81 and of CSM91 show that at small radii the
rotation velocity is very similar to that estimated from the \HI\ in
the outer regions, so the rotation curve of IC~5063 appears to be
quite flat. Assuming a flat rotation curve, we find for the
mass-to-light ratio at 1.3 $R_{\rm eff}$ (i.e.~the radius out to which
velocities can be measured from the optical data) a value of $M/L_B
\sim 5$.  We have not taken into account the fact that the inclination
of the inner disk could be slightly different from that of the outer
disk, but this can have only a small effect on our estimates of $M/L$,
since this inclination is in any case relatively large.  We have also
neglected possible effects due to non-circular orbits related to a
possibly triaxial potential in IC~5063, but it appears that the
increase of $M/L_B$ with a factor of almost 3 does point to the
presence of a halo of dark matter in IC~5063. The observed values for
$M/L_B$ in IC~5063 quite accurately follow the trend of $M/L$ with
radius noted for other early-type galaxies by Bertola et al.\ (1993).

\section{CONCLUSIONS}

We have presented radio continuum and \HI\ observations of the Seyfert
2 galaxy IC~5063 to investigate the ionization mechanism in the NLR as
well as the kinematics of the gas in different phases.  The high
resolution of our 8~GHz image reveals a linear structure that presents
strong morphological association with the NLR and is aligned with the
optical dust lane.

Very broad ($\sim$700 \kms) \HI\ absorption is observed against the
strong continuum source, indicating a fast net outflow.  This outflow,
together with the morphological correspondence between the radio
emission and the NLR, represent a strong argument for the presence of
shocks resulting from the interaction between the radio plasma and the
interstellar medium in the inner few kpc. Nevertheless, the energy
flux estimated from the radio emission is not enough to power the
emission of the optical lines. Although the morphology and the broad
\HI\ absorption suggest that shocks play a role at some level locally,
UV photons from the active nucleus are necessary to explain the
ionization of the NLR.

The main structure of the \HI\ emission is a warped disk.  Shadowing
effects from this disk are likely to be important in explaining the
morphology of the EELR.  Moreover, the neutral gas does not appear to
be kinematically disturbed outside the inner few kpc -- the kinematics
of the \HI\ connects smoothly that of the ionized gas in the center.
This indicates that both the ionized and neutral gas components are
just different phases of the same structure and this argues in favour
of an anisotropic radiation field responsible of the EELR morphology.

From the kinematics of the ionized and of the neutral gas, we find
evidence for a dark halo in IC~5063, with very similar properties as
observed in some other early-type galaxies.

\acknowledgments 

This research has made use of the NASA/IPAC Extragalactic Database
(NED) which is operated by the Jet Propulsion Laboratory, Caltech,
under contract with NASA. The Digitized Sky Survey (DSS) was produced
by the Space Telescope Science Institute (STScI). The DSS is based on
photographic data from the Oschin Schimdt Telescope, which is operated
by the California Institute of Technology and Palomar Observatory, and
the UK Schmidt Telescope, which is operated by the Royal Observatory
Edinburgh, the UK Science and Engineering Research Council and the
Anglo-Australian Observatory. ZT acknowledges support from NASA LTSA
grant NAGW-4443 to the Johns Hopkins University.

\newpage 


\centerline{\bf FIGURE CAPTIONS}

\figcaption[Morganti.fig1.ps]
	{The ATCA 8~GHz radio continuum image. The contour levels 
	are 0.75, 1.5, 3, 6, 12, 24, 48, 96\% of the peak value of 
	163 mJy beam$^{-1}$. Positions of the 21 cm continuum peak 
	and of the {\sc H i} absorption are marked with crosses with 
	sizes representing the 1$\sigma$ error bar. \label{Fig:3cm} }

\figcaption[Morganti.fig2.ps]
	{The ATCA 1.4~GHz radio continuum image superimposed onto 
	the DSS image. The contour levels are 2.5 mJy beam$^{-1}$ to 
	1.16 Jy in steps of factor 1.5. \label{Fig:21cm-opt} }

\figcaption[Morganti.fig3.ps]
	{The [{\sc O iii}] $\lambda$5007 image (contours) superimposed 
	onto the 8~GHz map (greysclage) with an enlargement of the 
	central region. \label{Fig:3cm-oiii} }

\figcaption[Morganti.fig4.ps]
	{Position-velocity slice taken along the major axis 
	(P.A. = 120$^\circ$). Contour levels: --1 to --16.16 mJy 
	beam$^{-1}$ in steps of 1.6 mJy beam$^{-1}$ (dashed lines) 
	and 1 to 5 mJy beam$^{-1}$ in steps of 1 mJy beam$^{-1}$ 
	(solid lines). \label{Fig:xv120} }

\figcaption[Morganti.fig5.ps]
	{The {\sc H i} absorption profile. \label{Fig:abs-prof} }

\figcaption[Morganti.fig6.ps]
	{The global {\sc H i} profile. \label{Fig:prof-hi} }

\figcaption[Morganti.fig7.ps]
	{Images from the individual velocity channels (plotted every 
	second channel). The contour levels are: $-$15, $-$8, $-$4, 
	$-$2, $-$1, 1, 1.5, 2, 3, 4, 5, 6, 7, 8, 9, 10  mJy beam$^{-1}$.
	\label{Fig:chan} }

\figcaption[Morganti.fig8.ps]
	{Total {\sc H i} intensity map of IC~5063 superimposed onto
	the optical image of the galaxy.  The ``hole'' in the center of 
	the {\sc H i} distribution (and coincident with the center of 
	the galaxy) represents the region of strong absorption against 
	the radio continuum. Contour levels: $1.0 \times 10^{20}$ to 
	$1.72 \times 10^{21}$ in steps of $1.8 \times 10^{20}$ atoms 
	cm$^{-1}$. \label{Fig:tothi-opt} }

\figcaption[Morganti.fig9.ps]
	{Total {\sc H i} intensity map of IC~5063 (contours as in 
	Fig.~8) superimposed onto the [{\sc O iii}] $\lambda$5007 
	image. \label{Fig:tothi-oiii} }

\figcaption[Morganti.fig10.ps]
	{The {\sc H i} velocity field of IC~5063. Contour levels range 
	from 3100 to 3700 in steps of 20 km s$^{-1}$. \label{Fig:vel} }

\figcaption[Morganti.fig11.ps]
	{The distribution of the high-excitation ionized gas as 
	outlined by the [{\sc O iii}] $\lambda$5007/H$\alpha$+[{\sc N ii}] 
	ratio map (white contours) superposed onto the $HST$ image. 
	The enlargement shows the 8~GHz map (green contours) and the 
	morphology of the central part of the $HST$ image. The 8~GHz 
	contours are drawn at 0.5, 1, 2, 4, 8, 16, 32, 64 \% of the peak 
	value of 163 mJy beam$^{-1}$. \label{Fig:ratio-hst} }

\clearpage





%
%
\begin{center}
{\sc Table 1.} {Properties of IC~5063}
\smallskip

\begin{tabular}{ll} \hline \hline
  $v_{\rm helio}$ (\kms)$^a$     & 3404  \\
  Distance (Mpc)$^b$             & 68    \\
  scale (kpc arcsec$^{-1}$)      & 0.32  \\
  $L_{B}$ (\Lsun)$^c$            & $4.7\times 10^{10}$ \\
  $M_{{\rm H_2}}$ (\Msun)$^d$    & $6.7\times 10^8$    \\
  $M_{{\rm HI}}$ (\Msun)$^e$     & $1.0\times 10^{10}$ \\
  $M_{{\rm HI}}$ (\Msun)$^a$     & $8.4\times 10^{9}$  \\
  $L_{{\rm FIR}}$ (\Lsun)$^d$    & $4.7\times 10^{10}$ \\
  $S_{60\mu {\rm m}}$ (Jy)$^f$   & 6.53 \\
  $S_{100\mu {\rm m}}$ (Jy)$^f$  & 3.92 \\
  $S_{\rm 1.4GHz}$ (Jy)$^a$      & 1.26 \\
  $P_{\rm 1.4GHz}$ (\whz)        & $6.3\times 10^{23}$  \\ \hline \hline
\end{tabular}
\smallskip

\parbox{5.5cm}{\indent References --- (a) this work; (b) assuming H$_0$ = 
50 \kms\ Mpc$^{-1}$; (c) RC3; (d) Wiklind et al.\ (1995); 
(e) DGW81; (f) Knapp et al.\ (1989)}

\end{center}

\bigskip

%
%
\begin{center}
{\sc Table 2.} {\HI\ Observations}
\smallskip

\begin{tabular}{lclccc} \hline \hline
Date & ATCA         & Min-Max & Bandwidth (MHz) & Frequency & Time \\
     &Configuration & Baseline (m) & /channels & (MHz) & (h) \\ \hline
1995 Sep    & 750D  &31$-$719(4469)$^*$ & 16/256  & 1406 & 12 \\
1995 Dec    & 6C    & 153$-$6000        & 16/256  & 1406 & 5  \\
1996 Apr    & 6A    & 337$-$5939        & 16/512  & 1406 & 12 \\
1996 May    & 1.5D  &107$-$1439(4439)$^*$& 16/512 & 1406 & 12 \\
1996 Jun    & 6D    & 77$-$5878         & 16/512  & 1406 & 12 \\ \hline\hline
\end{tabular}
\end{center}

\centerline{$^*$ the longest baseline length using also the 6 km antenna 
is given in parenthesis.}

\bigskip

%
%
\begin{center}
{\sc Table 3.} {Instrumental Parameters of the \HI\ Observations}
\smallskip

\begin{tabular}{lc} \hline \hline
Field Center  (J2000.0) & 
$20^{\rm h}52^{\rm m}02^{\rm s}.0$~~ $-57^\circ 04' 09''$ \\
Synthesized beam (natural weighting) & $18\farcs2 \times 16\farcs6$, 
P.A.=$-$39.7$^\circ$    \\
Synthesized beam (uniform weighting) & $8\farcs2 \times 6\farcs7$, 
P.A.=$-69^\circ$   \\
Velocity of the band center (\kms)     &   3290   \\
Velocity resolution (\kms)             &   26     \\
rms noise in channel maps (\mJybeam)   &   0.53   \\ \hline \hline
\end{tabular}
\end{center}

\end{document}